\documentclass[%
	aps,prl,amsmath,amssymb,floatfix,twocolumn,superscriptaddress]%
	{revtex4-1}
\usepackage{multirow}
\usepackage{graphicx}
\usepackage{subfigure}
\usepackage{color}
\usepackage{hyperref}
\usepackage{bm}

\usepackage{amsfonts, relsize, color}
\usepackage{graphics}
\usepackage{graphicx}
\usepackage{subfigure}
\usepackage{hyperref}
\usepackage{color}

\newcommand{\Mmin}{\ensuremath{M_{\mathrm{min}}}}

\begin{document}
\title{First-order quantum phase transition in three-dimensional topological band insulators}

\author{Vladimir Juri\v ci\' c}
\affiliation{Nordita,  Center for Quantum Materials,  KTH Royal
Institute of Technology and Stockholm University, Roslagstullsbacken 23,
10691 Stockholm,  Sweden}

\author{D.~S.~L.~Abergel}
\affiliation{Nordita,  Center for Quantum Materials,  KTH Royal
Institute of Technology and Stockholm University, Roslagstullsbacken 23,
10691 Stockholm,  Sweden}

\author{A.~V.~Balatsky}
\affiliation{Nordita,  Center for Quantum Materials,  KTH Royal
Institute of Technology and Stockholm University, Roslagstullsbacken 23,
10691 Stockholm,  Sweden}
\affiliation{Institute for Materials Science, Los Alamos National Laboratory, Los Alamos, NM 87545, USA}

\date{\today}

\begin{abstract}
Topological states of matter are characterized by global topological
invariants which change their value across a topological quantum phase
transition.
It is commonly assumed that the transition between topologically
distinct noninteracting gapped phases of fermions is necessarily
accompanied by the closing of the band gap as long as the symmetries of
the system are maintained. 
We show that such a quantum phase transition is possible without closing
the gap in the case of a three-dimensional topological band insulator.
We demonstrate this by calculating the free
energy of the minimal model for a topological insulator, the
Bernevig-Hughes-Zhang model, and show that as the band curvature
continuously varies, a jump between the band gap minima corresponding to
the topologically trivial and nontrivial insulators
occurs. Therefore, this first order phase transition is a generic
feature of three-dimensional topological band insulators.
For a certain parameter range we predict a re-entrant topological phase
transition. We discuss our findings in connection with the recent
experimental observation of a discontinuous topological phase transition
in a family of topological crystalline insulators.
\end{abstract}

\maketitle

\emph{Introduction}.
The study of topological states of matter has emerged as one of the most
active topics in condensed matter physics since the theoretical
prediction
\cite{kane-mele,kane-mele1,BHZ-science2006,fu-kane2007,moore-balents}
and the experimental discovery of two- and three-dimensional
topological band insulators (TBIs)
\cite{molenkamp2007,hasan2008,hasan2008-1,hasan2008-2}.  Topological
states of matter are characterized by a global order
parameter that represents the topological invariant of the underlying
noninteracting Hamiltonian, which distinguishes these states from the
ordinary Landau-like states characterized by a local order
parameter \cite{hasan-kane,qi-zhang}. In particular, topological
insulators in both two and three dimensions (3D) are characterized by a
$\mathbb{Z}_2$ topological invariant \cite{kane-mele,moore-balents}.
From the point of view of the bulk band structure, this topological
invariant counts the parity of the number of the band inversions
occurring at the high symmetry points in the Brillouin zone. Upon the
transition from a topological to a topologically trivial phase these
topological invariants change.

Until recently, the topological phase transition (TPT) between
trivial and nontrivial insulators was believed to be continuous
(\textit{i.e.} of second order) and accompanied by a closing of the band
gap which occurs at the quantum-critical point separating the two
phases.
The argument for this is based on the underlying assumption that the
band structure parameters continuously evolve with the external
parameters, such as doping and pressure.
Hence, band inversions could only occur if the band gap continuously
closes and reopens across a TPT.
Analysis of the thermodynamics appears to reveal universal behavior,
depending on the dimensionality of the model \cite{moraissmith}.
However, recent experiments \cite{tjernbergPRB} on topological
crystalline insulators \cite{Dziawa2012,Tanaka2012,Hasan2012} appear to
show a discontinuous (\textit{i.e.} first order) TPT in Sn-doped
PbSe. If these experiments are accurate, then a fundamental
change in our theoretical understanding of TPTs
will be required in order to explain them.
Theoretical work has recently predicted a discontinuous topological
quantum phase transitions driven by electron--electron interactions in
topological insulators
\cite{trauzettel,bitan-pallab,trauzettel1,troyer}, but in this
Letter we address the much more fundamental question of whether first
order TPTs can be explained within a noninteracting theory.

Using thermodynamic considerations, we establish that a first order TPT
driven by a band structure parameter (such as the band curvature) or by
changes in temperature, can occur even in \emph{noninteracting}
topological insulators.
We compute the free energy of the Bernevig-Hughes-Zhang (BHZ) model
\cite{BHZ-science2006}, which provides a minimal low energy description
of a topological insulator in 3D \cite{zhang-natphys2009}.
We find that as the band curvature continuously evolves, a jump between
the minima in the free energy corresponding to a topological and a
trivial insulator occurs at a critical value of the band curvature.
The magnitude of the band gap is comparable but not equal on the two sides
of the transition.
On the other hand, we find that the analogous transition in two
dimensions is always second order.
Our result therefore shows that the first order topological
quantum phase transitions are a generic feature of 3D topological
insulators.
At finite temperature the TPT remains discontinuous.
In a certain regime of parameters, we find a re-entrant TPT
as a function of temperature.
Our results qualitatively agree with the existing experiments.

\emph{Model}.
The continuum Hamiltonian of the BHZ model in three spatial dimensions
with two orbitals and two spins can be written in momentum space as
$H=\int\frac{d^3{\bf k}}{(2\pi)^3}\Psi^\dagger({\bf k}) H({\bf k})
\Psi({\bf k})$ where \cite{BHZ-science2006,zhang-natphys2009}
\begin{equation}\label{eq:BHZ}
	H({\bf k})=v{\bm \Gamma}\cdot{\bf k} +(M-B k^2)\Gamma_0,
\end{equation}
and spinor $\Psi^\dagger_{\bf k} = [u^\dagger_{{\bf k},\uparrow},
v^\dagger_{{\bf k},\uparrow}, u^\dagger_{{\bf k},\downarrow},
v^\dagger_{{\bf k},\downarrow}]$, with $u,v$
denoting annihilation operators of the low-energy valence and conduction
electron bands with spin projections $\sigma=\uparrow,\downarrow$ and
momentum ${\bf k}$; a momentum cutoff $\Lambda$ is assumed and we use
natural units $\hbar=k_B=e=1$.
The $\Gamma$ matrices form a set of mutually anticommuting Hermitian
matrices, $\{\Gamma_\mu,\Gamma_\nu\}=2\delta_{\mu\nu}$,
$\mu,\nu=0,1,2,3$. The parameter $v$ is the Fermi velocity, which is
irrelevant for the topological considerations, and we thus set it to
unity. The parameters $M$ and $B$ are the band gap and the band
curvature, respectively, and determine the $\mathbb{Z}_2$ topological
index $\nu$ of a phase as $(-1)^\nu=-{\rm sign}(MB)$. If $M$ and $B$
have the same sign, this corresponds to the topological phase, while
otherwise a phase is trivial. Notice that the band
curvature term proportional to the square of the momentum, has to be
included in the effective Hamiltonian for the topological invariant to
be defined.
This Hamiltonian describes a time-reversal symmetric insulator
close to the TPT between a trivial and a
topological phase. It can be thought
of as a minimal continuum Hamiltonian that accounts for the
$\mathbb{Z}_2$ topological invariant, and  can be obtained from the
lattice tight binding models after expanding the Hamiltonian about the
band gap minimum.  It was originally introduced to describe
two-dimensional
quantum spin Hall insulator state realized in HgTe/CdTe quantum wells
\cite{BHZ-science2006}, and its 3D version provides an effective
description of Bi-based TBIs \cite{zhang-natphys2009}. We here use the
simplest version of the BHZ Hamiltonian where full 3D rotational
symmetry is present
\cite{felser-bbo,slager-natphys2013}. This Hamiltonian is in general
anisotropic, but the anisotropies are irrelevant for the topological
purposes and we thus do not include them.

\emph{Discontinuous topological quantum phase transition}.
To study the TPT between a topological and a trivial phase, we compute
the free energy density of the BHZ model at temperature $T=0$
\begin{equation}
	F_0=\int\frac{d^3{\bf k}}{(2\pi)^3}\varepsilon({\bf k}),
	\label{eq:Fzero}
\end{equation}
with
\begin{equation}\label{eq:dispersion}
	\varepsilon({\bf k})=\sqrt{M^2+(1-2MB)k^2+B^2k^4},
\end{equation}
as the dispersion corresponding to the Hamiltonian (\ref{eq:BHZ}).
We consider the thermodynamic limit and so neglect the contribution from
the surface states since it is proportional to the surface to bulk
volume ratio which tends to zero in this limit.
Using the rotational symmetry and introducing the substitution $x=k^2$,
we obtain
\begin{equation}
	\label{eq:FMB}
	F_0(M,B) = \frac{1}{2\pi^2}
	\int_0^{\Lambda^2} dx\sqrt{x[M^2+(1-2MB)x+B^2x^2]}.
\end{equation}
This integral cannot be calculated in the closed form so we expand the
integrand to arbitrary order in the parameter $B$, and then compute it.
Finally, we extract the cutoff-independent part of the integral, which
therefore only depends on the parameters $M$ and $B$.
The result to the $n^{\rm th}$ order in $B$ is
\begin{equation}\label{eq:free-energyT0}
	F_0^{(n)}(M,B) = \frac{M^4}{D_n}
	\left[ A_n(MB)+ C_n(MB)\log M^2\right],
\end{equation}
with $A_n(x)=\sum_{k=0}^n a_k x^k$ and $C_n(x)=\sum_{k=0}^n c_k x^k$,
where $a_k$ and $c_k$ are real coefficients, and $D_n$ is also real; for
details, see the Supplemental Material \cite{SM}.
Notice that the free energy satisfies $F_0(M,B)=F_0(-M,-B)$.
The term
containing the logarithm of the band gap arises from the fact that Dirac
fermions live in three dimensions where hyperscaling is violated
\cite{sachdev}, and together with odd powers of the parameters
$M$ and $B$ in the free energy (\ref{eq:free-energyT0}), is responsible
for the discontinuous character of the TPT, which we discuss next.
On the other hand, repeating this calculation in two spatial dimensions
shows that the free energy does not contain the logarithm and the
topological transition is continuous and accompanied by a closing of the
band gap.

\begin{figure}[t]
\includegraphics[]{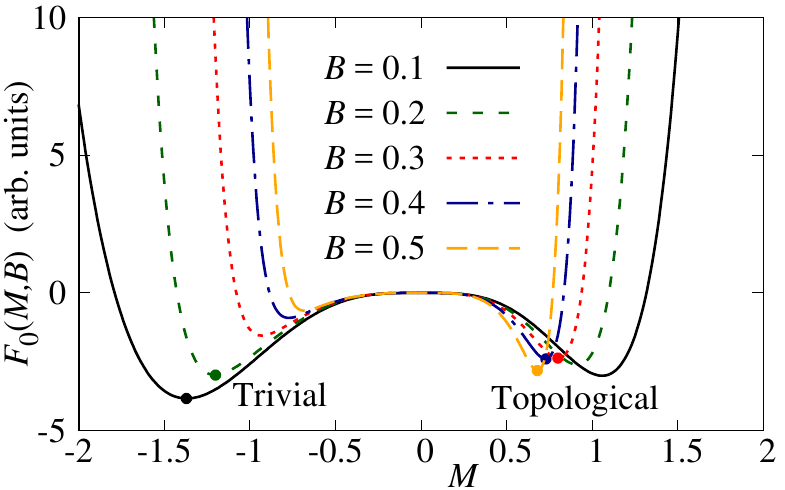}
\caption{(Color online) Zero temperature free energy as a function of
the  band gap for different values of the band curvature.
The green and red curves represent the free energy right before and
after the transition from the topological to the trivial phase at the
critical $B_{c}(T=0)\simeq0.23$.
Notice that the band gaps corresponding to the global minima in the
topological and the trivial phase across the transition are similar in
size, but not precisely equal. Global band gap minima are marked by a
dot for each curve.}
\label{fig:free-energy-diffB}
\end{figure}

\begin{figure}[t]
\includegraphics[]{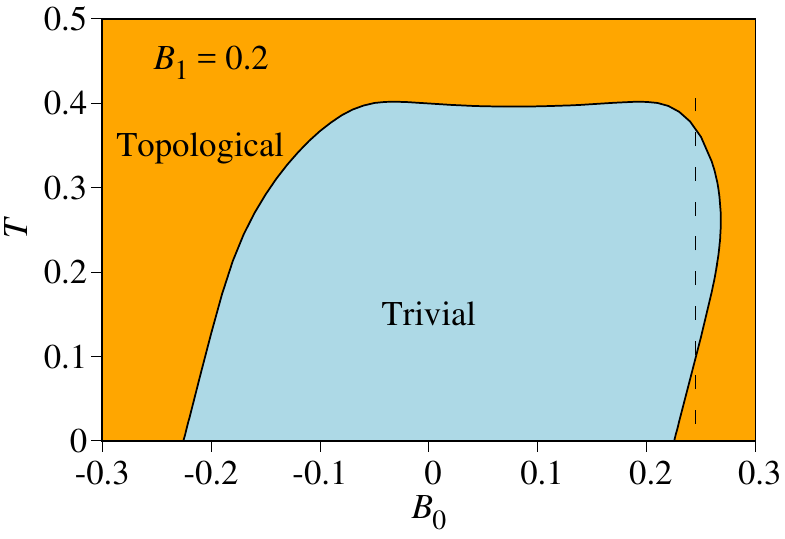}
\caption{(Color online) Finite temperature phase diagram of the BHZ
model as a function of the zero temperature band curvature $B_0$.
The trivial phase remains stable at finite temperature up to a critical
value of the zero temperature band curvature.
Notice that there is a region where a re-entrant behavior in the
topological phase through the trivial phase is possible, highlighted by
the vertical dashed black line.
For larger absolute values of  $B_0$ topological phase is stable at any
temperature. The topological and trivial phases are separated by a
first-order phase transition. }
\label{fig:phasediagram}
\end{figure}

We now study the above zero temperature free energy of the BHZ model as
function of the band curvature. In Fig.~\ref{fig:free-energy-diffB} we
plot the free energy to the eighth order in $B$ for different positive
values of this parameter.
The free energy has two local minima at values of the
gap parameter asymmetrically located with respect to $M=0$ with the
global minimum at $\Mmin>0$ ($\Mmin<0$) corresponding to the topological
(trivial) phase.
Most importantly, as the band curvature increases, the global minimum in
the free energy \emph{discontinuously} changes from a positive value to
a negative one at the critical value $B_{c}(T=0)\simeq0.23$,
as shown in Figs.~\ref{fig:free-energy-diffB} and
\ref{fig:phasediagram}.
The green and red curves in Fig.~\ref{fig:free-energy-diffB} correspond
to the free energy right before and after the transition. Notice that
the absolute values of the  band gap minima on the two sides of the
transition are comparable in size.

\emph{TPT at finite temperature}.
The free energy at finite temperature has an additional contribution
coming from the entropy, $F(M,B,T)=F_0(M,B)-TS(M,B,T)$, with
\begin{equation}\label{eq:entropypart}
	S(M,B,T) = \int\frac{d^3{\bf k}}{(2\pi)^3}
	\log\left(1+e^{-\frac{\varepsilon({\bf k})}{T}}\right),
\end{equation}
with the dispersion given by Eq.~(\ref{eq:dispersion}), and is also even
under the combined operation $M\rightarrow -M$ and $B\rightarrow -B$.
Furthermore, it has been demonstrated by recent numerical studies of
topological insulator crystals \cite{vanderbilt, louie} that
the band curvature parameter $B$ may also depend on temperature.
Therefore, we generalise $B$ to the form $B(T)=B_0-B_1 T$, where $B_1$
is a positive parameter which describes the response of the lattice to
finite temperature. The negative sign is chosen since we assume that the
lattice expands with increasing temperature, \textit{i.e.} it has a positive thermal
expansion coefficient, indicating that the
effective lattice constant increases, hence weakening the confining
potential which generates the band curvature and so $B$ decreases.
We choose linear temperature dependence of the band curvature for
simplicity to demonstrate the effect of temperature on the TPT.
However, only its monotonic behavior
with temperature is important in this regard. At finite temperature, the
full $B(T)$ is substituted into
Eqs.~\eqref{eq:Fzero}--\eqref{eq:entropypart} to compute the free
energy.

The finite temperature phase diagram is shown in
Fig.~\ref{fig:phasediagram}.
Most importantly, the discontinuous character of the TPT
remains unchanged at finite temperature, so that all lines in
the phase diagram correspond to first order phase transitions.
A system in the trivial phase at $T=0$ always undergoes a first order
transition into the topological phase at high temperature as a result of
the entropic contribution to the free energy.
However, the low temperature regime is more subtle, and depends on the
competition between the entropic contribution and the temperature
dependence of the band curvature.

Let us first address the situation $B_0>0$, which corresponds to TBIs
with a positive thermal expansion coefficient,
shown in the right half
of Fig.~\ref{fig:phasediagram}.
We observe that as the value of the zero temperature band
curvature ($B_0$) increases, the critical temperature ($T_c$) for the
phase transition from the trivial to the topological phase increases.
This occurs because as temperature increases, band curvature
decreases and therefore the transition to the topological phase should
occur at $B_0>B_{c}(T=0)$, \textit{i.e.}, at a value of $B_0$ higher
than the critical one at $T=0$. 
By neglecting the entropy contribution in the free energy
(\ref{eq:entropypart}), we can estimate critical temperature $T_c$ from
the value of the critical band curvature at zero temperature,
$B_{c}(T=0)\approx B_0-B_1 T_c$, yielding
$T_c\approx[B_0-B_{c}(T=0)]/B_1$.
We see in Fig.~\ref{fig:phasediagram} that this trend indeed holds when
$T_c$ is low enough and the effects from the band curvature dominate.
At higher temperatures, thermal fluctuations take over and bend the
phase boundary to favor the topological phase. 
As a result, there is a region in the phase diagram $0.23 \lesssim
B_0 \lesssim 0.28$ where, as temperature increases
from zero, the system begins in the topological phase, undergoes a
first-order transition to the trivial phase and then re-enters the
topological phase at some higher temperature, as marked by a dashed
black line in Fig.~\ref{fig:phasediagram}.
The size of the region where this re-entrant behavior takes
place scales with the slope of the band curvature at low
temperature. 
The case $B_0<0$ describes TBIs with a negative thermal expansion
coefficient since $F(M,B) = F(-M,-B)$ and is shown in the left half of
Fig.~\ref{fig:phasediagram}. The previous argument suggests that the
$T_c$ decreases linearly with $|B_0|$, which is seen in the
low temperature regime. 
At higher temperature, we observe that $T_c$ crosses over from linear to
nonlinear (quadratic) dependence on $|B_0|$ as the thermal fluctuations
start to become important.

\begin{figure}[b]
\includegraphics[]{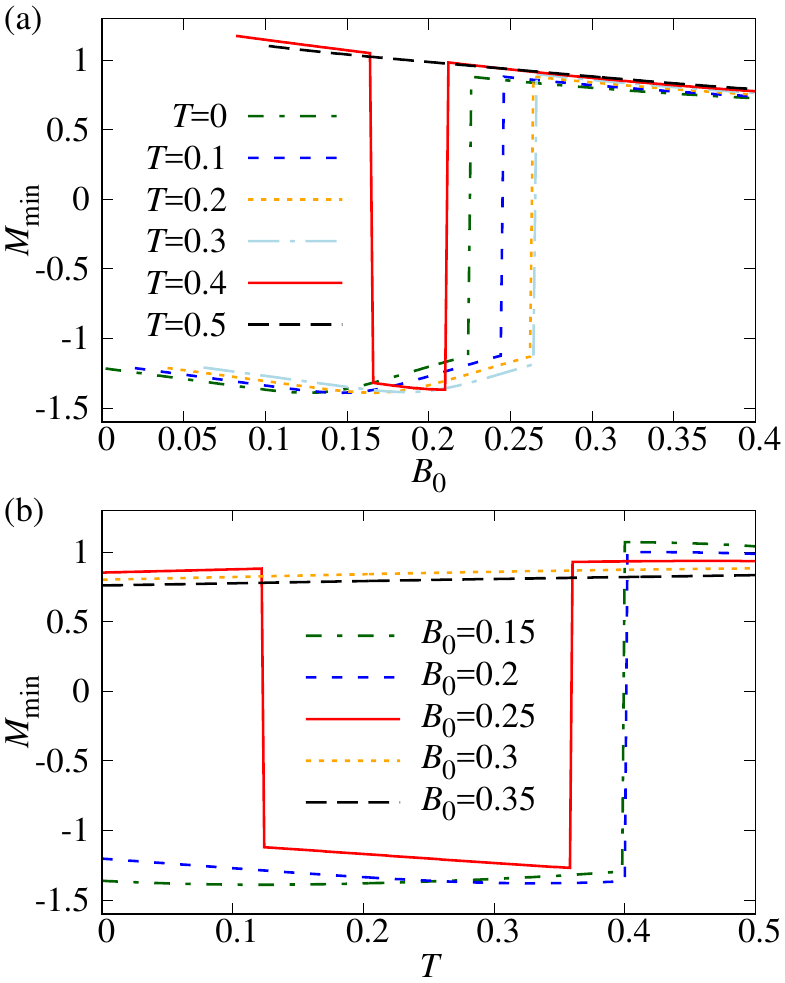}
\caption{(Color online)  Evolution of the global band gap minimum
$\Mmin$ as a function of $B_0>0$  corresponding to the phase diagram in
Fig.\ \ref{fig:phasediagram}. $\Mmin$ vs. $B_0$ at different
temperatures; (b) $\Mmin$ vs. $T$ for different $B_0$.
Notice that in both panels the magnitude of the band gap is comparable
across the transition.  The red curve in both panels corresponds to the
re-entrant TPT. }
\label{fig:gap-evolution}
\end{figure}

The absence of the band gap closing is shown in
Fig.~\ref{fig:gap-evolution} where we plot $M_\mathrm{min}$, the value
of the band gap at the global minimum of the free energy as a function
of $B_0>0$ and $T$.
In panel (a) we observe that as the temperature increases for
$0<T\lesssim0.3$, the critical value of $B_0$ for the transition into
the topological phase, $B_{0c}(T)$, increases. This can be understood,
neglecting the entropy contribution (low temperature), by noticing that
the total band curvature at the transition is equal to the critical one
at $T=0$, and therefore $B_{0c}(T)=B_{c}(T=0)+B_1 T$. However, at
higher temperature, the effects of thermal fluctuations start to be
important, which leads to a stable topological phase, after a narrow
region of parameters where a re-entrant behavior is possible. In Fig.
\ref{fig:gap-evolution}(b), we plot the band gap as a function of
temperature, from which we can infer that at values of
$B_0<B_{c}(T=0)$, trivial phase is stable up to $T\sim0.4$. When
$B_0\approx B_{c}(T=0)$, we observe the re-entrant behavior described
previously, where the system undergoes two discontinuous topological
transitions as the temperature increases. Finally, for even higher
values of $B_0$, the system is always topological, since already at
$T=0$ the system in this regime is topological and thermal fluctuations
further stabilize it.

\emph{Comparison with experiments}.
To demonstrate the utility of this theory, we make a qualitative
comparison of our findings with recent experiments in
Ref.~\onlinecite{tjernbergPRB}, where a phase transition
without an associated closing of the band gap has been reported in
Pb$_{1-x}$Sn$_x$Se, as a function of both doping and temperature.
This alloy has a face-centered cubic structure on both sides of the TPT,
up to a doping level of $x\approx 0.4$ \cite{Dziawa2012}.
We first consider the doping dependence.
For doping $x\lesssim x_c=0.17$ and at low temperature, the experiments
show that this compound is topologically trivial, while for $x>x_c$ it
features band inversions at four inequivalent $L$ points in the
Brillouin zone and is therefore a topological crystalline insulator
\cite{fu-natcomm2012,Dziawa2012}. The Hamiltonian in Eq.~\eqref{eq:BHZ}
describes the band inversion at each of these $L$ points.
The band curvature is linked to the doping $x$ in the Pb$_{1-x}$Sn$_x$Se
materials.
This is due to the Sn doping decreasing the effective lattice constant
$a$ and hence stiffening the confining potential which generates the
band curvature \cite{fu-natcomm2012}.
Therefore, $B_0\propto x$, and our theory gives precisely the behavior
observed in the experiment.

We now move on to the temperature dependence.
When the material is in the topological phase at low temperature, the
experiments show that raising the temperature at a fixed doping drives a
first-order TPT to the trivial phase. We
propose that the relevant $B_0$ parameter for Pb$_{1-x}$Sn$_x$Se
at maximum doping is in the range $0.23<B_0<0.28$, with $B_1>0$ so that
this transition is the initial (low temperature) part of the re-entrant
behavior we mark in Fig.~\ref{fig:phasediagram} with a dashed line.
Indeed, for $x>x_c$, the experimental results show that as the doping
increases, the critical temperature also increases, which is in
agreement with our results in Fig.~\ref{fig:phasediagram}.
Moreover, as the temperature increases, the doping region with the
topological phase shrinks up to a temperature above which the trivial
phase is realized, also consistent with the experimental results.
We suspect that the range of temperatures accessed in
Ref.~\onlinecite{tjernbergPRB} does not go high enough to see the second
part of the re-entrant behavior which manifests as a phase transition
back to the topological phase. Indeed, the $T_c$ of the first transition
at $T_c\approx 250\mathrm{K}$ is already close to the maximum
temperature of $T=300\mathrm{K}$ reported in the experiment.

When $x<x_c$, the material is in the trivial phase at low temperature.
The experiments show that it never undergoes a phase transition to the
topological phase up to approximately $T=300\mathrm{K}$. Again, we
suggest that a topological phase may reappear due to its stabilization
by the entropic contribution to the free energy at higher temperatures.
Finally, we point out that our calculation gives sizes of the band gaps
in the trivial and the topological phases which are comparable in
magnitude immediately on either side of the TPT, also consistent with
the experimental findings.

\emph{Conclusions}. Here we have established the first-order nature of the
topological quantum phase transition in \emph{noninteracting} 3D TBIs.
We also show that at finite temperature the transition remains
discontinuous. Gross features of the finite temperature phase diagram
depend on the behavior of the band curvature with temperature.
We find that there is a regime of parameters for which our theoretical
findings qualitatively agree with experiments.
However, we point out that the temperature and doping dependence of the
band curvature for topological insulators will vary between different
materials \cite{vanderbilt, louie}, with phonons possibly playing an
important role   \cite{garate,garate1}.
We therefore hope that our findings will motivate {\it ab initio}
calculations to ascertain these details.
We also show that it is possible to infer the temperature dependence of
the band curvature from the experimental behavior of $T_c$ as a function
of $B_0$ near to $B_c(T=0)$, since in this regime the entropy
contribution due to thermal fluctuations is expected to be subdominant.
Therefore, this work motivates further experiments to probe the nature
of the quantum and finite temperature TPTs in insulators, as well as in
superconductors.
Indeed, a first-order TPT has been also found in Sn-doped PbTe, with an
even more pronounced discontinuity than in the case of PbSe
\cite{Assaf-privatecomm}.

We are grateful to O.~Tjernberg, and B.~Roy for useful discussions, and
B.~Trauzettel for constructive critique.
This work was supported by Nordita, and KAW.
The work of AVB was supported by US DOE E3B7.

\end{document}